\documentclass{article}
\usepackage{ijcai24}

\usepackage{todonotes}

\usepackage{times}
\usepackage{soul}
\usepackage{url}
\usepackage[hidelinks]{hyperref}
\usepackage[utf8]{inputenc}
\usepackage[small]{caption}
\usepackage{graphicx}
\usepackage{amsmath, amssymb}
\usepackage{amsthm}
\usepackage{booktabs}
\usepackage{algorithm}
\usepackage{algorithmic}
\usepackage[switch]{lineno}
\usepackage{adjustbox} 
\usepackage{xcolor}
\usepackage{colortbl}

\newcommand{\newterm}[1]{\textbf{\textit{#1}}}

\usepackage{xcolor}

\title{Policy Space Response Oracles: A Survey}

\author{%
Ariyan Bighashdel*$^{,12}$\and
Yongzhao Wang*$^{,\dagger,3}$\and
Stephen McAleer$^{4}$\and 
Rahul Savani$^{35}$\\\And 
Frans A. Oliehoek$^{1}$\\
\affiliations
{*Equal contributions, $\dagger$ Corresponding author\\
$^1$Delft University of Technology, NL\\ $^2$Eindhoven University of Technology, NL\\ 
$^3$The Alan Turing Institute, UK\\  $^4$Carnegie Mellon University, USA\\
$^5$University of Liverpool, UK
}
\emails
{
\{a.bighashdel,f.a.oliehoek\}@tudelft.nl,
yongzhao.wang@turing.ac.uk,
smcaleer@cs.cmu.edu,
rahul.savani@liverpool.ac.uk
}
}

\begin{document}

\maketitle

\begin{abstract}

Game theory provides a mathematical way to study the interaction between multiple decision makers. However, classical game-theoretic analysis is limited in scalability due to the large number of strategies, precluding direct application to more complex scenarios.
This survey provides a comprehensive overview of a 
framework for large games, known as Policy Space Response Oracles (PSRO), which holds promise to improve scalability by focusing attention on sufficient subsets of strategies. 
We first motivate PSRO and provide historical context.
%
We then focus on the strategy exploration problem for PSRO: the challenge of assembling effective subsets of strategies
that still represent the original game well with minimum computational cost. 
%
We survey current research directions for enhancing the efficiency of PSRO,
and explore the applications of PSRO across various domains.
We conclude by discussing open questions and future research.

\end{abstract}

 
\section{Introduction}


In recent decades, the exploration of multiagent systems has been a central focus in Artificial Intelligence (AI) research. 
A multiagent system, often referred to as a \newterm{game}, comprises multiple decision-making agents that interact within a shared environment. 
To understand strategic behavior among these agents -- where  the optimal behavior of one agent depends on the behavior of others --
game theory provides a mathematical framework that defines behavioral stability through solution concepts like the Nash equilibrium (NE).
To identify such solutions, various \newterm{equilibrium computation} approaches have been developed~\cite{von2002computing,gambit23}, either
to enumerate all equilibria (see the work by Avis~et~al.~\shortcite{avis2010enumeration} for bimatrix games), or to find a single \newterm{sample equilibrium} (e.g., with the Lemke-Howson algorithm for a bimatrix game, or Linear Programming for a zero-sum matrix game~\cite{von2002computing}). 
 
As the size of the game (i.e., the number of players and strategies) grows, the computational feasibility of enumeration diminishes, and one
tends to focus on finding a sample equilibrium.
In the special case of two-player zero-sum games (i.e., settings where two players strictly compete),  
a sample equilibrium already provides valuable insights: it 
reveals the game's unique 
\emph{value}, which is the payoff that the first player can guarantee to obtain by playing a sufficiently strong (equilibrium) strategy \emph{irrespective} of the strategy of the other player.
However, even in zero-sum settings, many games that arise from practical applications are too large, and computing a sample equilibrium 
(even with polynomial-time methods for the resulting linear program) is infeasible.
%
It is these huge games that are our primary focus in this survey.

As an alternative to traditional equilibrium computation methods, to reason about such huge games, a wide range of \newterm{learning} methods have been applied.
Applying learning methods to games is known as \newterm{multiagent learning} \cite{shoham2007if}, with one of 
the most prominent approaches being multiagent Reinforcement Learning (RL)~\cite{marl-book}.
Compared to traditional methods, learning methods reduce the need to represent the entire game and create intelligent agents by exploring the game interactively.
While learning approaches have significantly contributed to the development of intelligent agents, they face many inherent challenges in games.
For example, independent learning across agents can render the environment \newterm{non-stationary}, which is a challenge for convergence as each individual learner faces
a potentially moving target \cite{tuyls2012multiagent}.
Another challenge is \newterm{non-transitivity} of a game, where there is not a clear notion of ``better'' strategy for an agent, and thus effective learning
requires the learning scheme to maintain a population of strategies for each agent.
Such non-transitivity exists ubiquitously in various games \cite{czarnecki2020real,sanjaya2022measuring,li2023jiangjun}.

Against this backdrop, the \newterm{Policy Space Response Oracles} (PSRO) framework~\cite{lanctot2017unified} emerged as a natural combination of traditional game-theoretic equilibrium computation with learning. 
In PSRO, a key concept is a \newterm{restricted game} with estimated payoffs\footnote{To be precise, normally a restricted game 
constrains the strategy space (relative to the full game), but the payoffs are ``exact'', rather than being estimates.
By contrast, in practice, PSRO implementations would typically estimate these payoffs through simulation (with stochasticity
coming potentially from the environment). A restricted game with estimated payoffs would normally be termed an \newterm{empirical game}~\cite{wellman2006methods} or \newterm{meta-game}~\cite{lanctot2017unified}. For simplicity, in this survey, we use the term ``restricted game'' throughout, regardless of whether payoffs are estimated or not.}, which acts as an approximation of the underlying full game.
Restricted games are induced from simulations run over combinations of a particular set of strategies.
This set of strategies is typically much smaller than the full game, and thus restricted games are feasible to analyze with traditional equilibrium computation methods.
PSRO iteratively expands the restricted game by introducing new strategies generated via learning, based on the analysis of the current restricted game.

%

As a general solver for large-scale games, PSRO has been successfully applied to a wide range of game types and diverse application domains, from mechanism design for sequential auctions~\cite{zhang2024computing} to robust RL~\cite{liang2023game}. 
Numerous PSRO variants have been developed, each tailored to leverage the specific characteristics of different underlying games. As a notable example of its success, algorithms inspired by PSRO have reached state-of-the-art performance in large-scale games such as Barrage Stratego~\cite{mcaleer2020pipeline}, and in StarCraft~\cite{vinyals2019grandmaster}, where they have convincingly outperformed human experts and prior AI systems.

%
While PSRO and its variants have been covered to some extent in existing multiagent learning surveys (e.g., \cite{yang2020overview,long2023survey,marl-book}), 
a dedicated survey on PSRO like this one has been lacking. 
In this survey, we reflect on the historical development of PSRO, which arose from different research communities, and 
we place PSRO within the space of game solving approaches. 
We then present the latest developments on PSRO, highlighting both current and future research directions.

\begin{figure*}[!hbpt]
	\centering
	\includegraphics[width=0.8\textwidth]{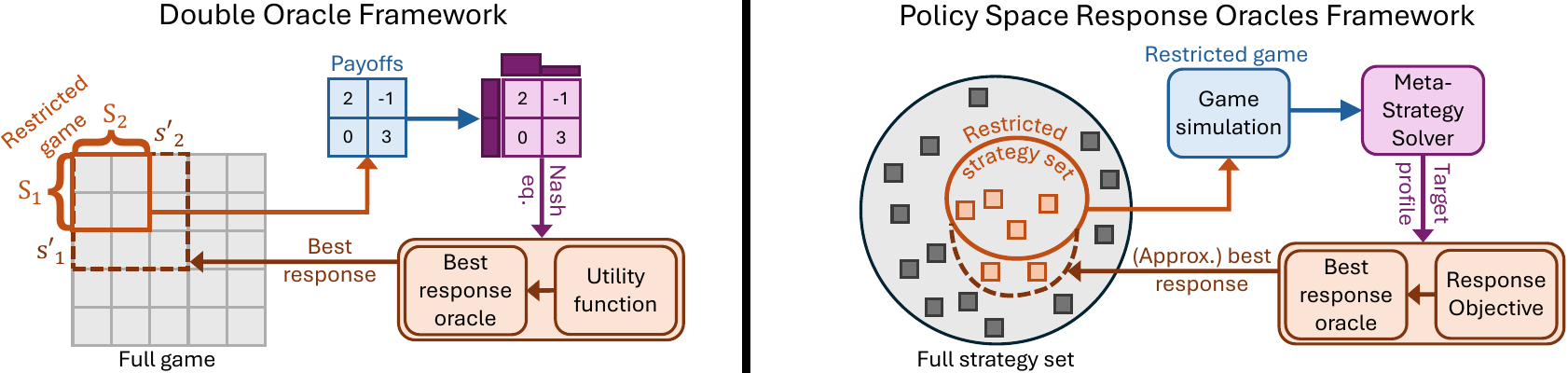}
 \vspace{-0.2cm}
	\caption[]{Illustration of the DO and PSRO frameworks respectively. The PSRO framework generalizes the DO framework by introducing MSSs, enabling best-response targets other than NE. Besides, PSRO accommodates various ROs and (approximate) best-response oracles.}
	\label{fig:psro}
 \vspace{-0.3cm}
\end{figure*}

\section{The PSRO Framework}
The PSRO framework incorporates a synthesis of ideas originating from distinct research communities. Within the \newterm{planning} community, McMahan et al.~\shortcite{mcmahan2003planning} laid the groundwork by formulating robust planning in Markov Decision Processes as a zero-sum game characterized by a vast strategy space. 
Drawing inspiration from Bender's decomposition in optimization~\cite{bnnobrs1962partitioning}, they introduced the \newterm{Double Oracle} (DO) algorithm. 
DO is an iterative algorithm for solving games with a finite number of strategies. 
DO maintains a restricted version of the full game and iteratively expands the restricted game by adding best responses to the current equilibrium.
When DO terminates, no player can deviate unilaterally to gain extra payoff and therefore the equilibrium in the current restricted game is an NE of the full game.
In finite games, DO is guaranteed to converge to an NE, though the restricted game could include all strategies of the full game in the worst case.
Moreover, under proper tie-breaking assumptions, Zhang and Shandholm~\shortcite{zhang2024exponential} showed that DO takes exponentially many iterations to converge in partially-observable stochastic games and extensive-form games.

Concurrently, similar methods were being developed from a different perspective in the \newterm{co-evolution} research community. 
Co-evolutionary methods evolve multiple populations of (different) species in parallel.
In this setting, there is no explicit fitness function, but the fitness of a population member depends on how well it interacts
with members of other populations.
Traditional challenges for co-evolution include the presence of intransitive cycles in traits, and the related problem of forgetting a trait that seems not useful at one point in the evolutionary process but later becomes useful again.
%
The community explored \newterm{memory mechanisms} and game formulations where the co-evolutionary process was set up to 
discover NE as mixtures of traits~\cite{angeline1993competitive,popovici2012coevolutionary}.
%
To overcome the mentioned obstacles of intransitive cycles and forgetting traits, various archival structures were devised to facilitate monotonic convergence towards a range of game-theoretic solution concepts. 
For example, the ``Nash memory" for symmetric games~\cite{ficici2003game} identifies equilibrium strategies within a discovered restricted game. This was extended to asymmetric games via the Parallel Nash Memory (PNM)~\cite{oliehoek2006parallel}, broadening the standard DO framework to accommodate alternative oracle types beyond the best response.


The PSRO framework~\cite{lanctot2017unified} was introduced in a paper titled ``A Unified
Game-Theoretic Approach to Multiagent Reinforcement  Learning'', bringing many of the mentioned ideas from the planning, multiagent RL, and co-evolution communities together in a unified framework, which also covers classical game-theoretic learning dynamics such as Fictitious Play~\cite{brown1951iterative}.
%
Technically, PSRO generalizes DO in three ways.
Firstly, PSRO introduces the concept of \newterm{meta-strategy solver} (MSS), which extracts a profile from the current restricted game as the next best-response target\footnote{For simplicity, we often use the same term for the solution
concept and the MSS that computes it. For example, we may say ``Nash equilibrium'' to mean the MSS that computes an NE.}.
This enables the best-response target to extend beyond NE, transforming PSRO into a versatile framework capable of generalizing various classical and modern game-theoretic algorithms.
Secondly, PSRO generalizes DO by allowing any form of (approximate) response oracles, including search, planning, and evolutionary algorithms etc (as used in PNM and other works~\cite{oliehoek2006parallel,li2021evolution}).
This generalization enables best-response computation in environments with a large number of states and actions. 
Thirdly, compared to DO, the payoffs of profiles in PSRO are estimated through simulation. 
See Figure~\ref{fig:psro} for a depiction of the whole framework, contrasted with DO.

We note that the PSRO framework can be thought of as an instance of \newterm{Empirical Game-Theoretic Analysis} (EGTA)~\cite{wellman2006methods}, which includes a broad set of methods that build and analyze restricted games based on simulation.
%
%
A comprehensive introduction to EGTA can be found in the survey by Wellman~et~al.~\shortcite{wellman2024empirical}.
As a concrete example of the connection between EGTA and PSRO, 
in an early EGTA work, Schvartzman and Wellman~\shortcite{Schvartzman09a} deployed tabular RL as a best-response oracle (at a time when deep RL did not exist yet) and NE of the restricted game as a best-response target for strategy generation (i.e. as the MSS).
%


\subsection{The Framework}
A normal-form (aka strategic-form) representation of the full game $\mathcal{G} = (N, (S_i), (u_i))$ is a tuple, where $N$ is a finite set of players, each with a non-empty set of strategies $S_i$ and a utility function $u_i: \Pi_{j \in N} S_j \rightarrow \mathbb{R}$. 
A restricted game $\hat{\mathcal{G}}_{S\downarrow X} = (N, (X_i), (\hat{u}_i))$ is a projection of the full game $\mathcal{G}$, with players choosing from restricted strategy sets $X_i \subseteq S_i$, allowing for utilities to be estimated via simulation. 


Figure~\ref{fig:psro} shows the special case of DO applied to a bimatrix game on the left, and the general PSRO framework on the right.
In PSRO, each player is initialized with a set of strategies $X_i$ and the utilities for profiles in the profile space $X$ are simulated, resulting in an initial restricted game $\mathcal{\hat{G}}_{S\downarrow X}$.
At each iteration of PSRO, an MSS designates a profile $\sigma \in \Delta X$ from the current restricted game $\mathcal{\hat{G}}_{S\downarrow X}$ as the next best-response target, where $\Delta$ represents the probability simplex over a set.
Then each player $i\in N$ independently computes (learns) a best response $s_i' \in S_i$ against its \newterm{response objective} (RO), which is a function of strategy profiles, denoted as $\mathit{RO}_i(\sigma)$.
In standard PSRO, the RO can be written as $\mathit{RO}_i(\sigma) = u_i(s_i', \sigma_{-i})$ and maximizing it over $s_i'$ gives player $i$ a best response against other players' strategies~$\sigma_{-i}$. 
During this procedure, the other players' strategies~$\sigma_{-i}$ are fixed, which renders the environment stationary for the learning player to compute their response.
Then the best response~$s_i'$ will be added to its strategy set $X_i$ in the restricted game.
This procedure repeats until a stopping criterion has been satisfied (e.g., a fixed number of iterations have been completed or the estimated regret of the restricted-game NE is below a threshold).

\subsection{Strategy Exploration in PSRO}\label{sec: SE in PSRO}
In essence, game-theoretic analysis in PSRO is performed by reasoning about restricted games.
A restricted game is expected to contain an effective subset of strategies for representation tractability yet still represent the full game well strategically \cite{balduzzi2018re}.
This challenge of restricted game construction with minimum computational cost (i.e., with the fewest strategies required) is described as the \newterm{strategy exploration} problem \cite{Jordan10sw}, which is the main research focus for developing PSRO methods. 
In PSRO, strategy exploration can be controlled by setting MSSs and ROs, which 
have a coupled impact;  we refer to the joint choice as an MSS-RO combination.

The performance of strategy exploration given a specific MSS-RO combination is normally monitored through the concept of \emph{regret}. 
The regret $\rho_i(\sigma)$ for a player $i$ in a strategy profile $\sigma$ is the difference between the player's payoff under $\sigma$ and the payoff they could have achieved by employing their best-response strategy. Formally, it is defined as $\rho_i(\sigma) = \max_{s'_i \in S_i}\; u_i(s'_i, \sigma_{-i}) - u_i(\sigma_i, \sigma_{-i})$. 
This measure reflects the maximal expected gain of player $i$ from unilaterally deviating from their current mixed strategy in $\sigma$ to an alternative strategy in $S_i$. 
In an NE, each player's strategy is a best response to the strategies of the others, which implies that no player can gain by unilaterally changing their strategy. 
Consequently, a profile is an NE if and only if its regret is zero for all players. 
Moreover, the stability of a profile $\sigma$ depends on the aggregation of regrets over players.
There are basically two natural ways to aggregate regret over players: the max over regrets and the sum of regrets.
Specifically, the sum of regrets of a strategy profile~$\sigma$ over players, denoted as $\rho(\sigma)=\sum_{i \in N} \rho_i(\sigma)$, is known as $\mathit{NashConv}(\sigma)$~\cite{lanctot2017unified}. $\mathit{NashConv}(\sigma)$ measures how far the strategy profile is from NE. 
In the context of two-player zero-sum games, $\mathit{NashConv}(\sigma)$ is often referred to as \emph{exploitability}, indicating the extent to which the strategy profile can be exploited by an adversary. 
The second way to aggregate is taking the max of regrets over players.
The max of regrets is more standard than NashConv in game theory, as it directly corresponds to the definition of $\epsilon$-NE (i.e., a profile within which no player can gain more than $\epsilon$ by unilateral deviation).

In practice, the computation of regrets requires an exact best-response oracle, which is achievable in small games through methods such as strategy enumeration or dynamic programming. 
However, in large games, computing an exact best response becomes impractical. 
In such cases, approximate best responses are employed, providing a lower bound on regrets. The accuracy of regret estimation improves with a higher-quality oracle. 
With limited computational resources, when we cannot find a ``better'' response for any player, Oliehoek et al.~\shortcite{oliehoek2019beyond} named the resulting profile a \newterm{resource-bounded NE}, with the interpretation: ``with these resources, we did not refute that this is an NE''.

\subsection{RL View of PSRO: Population-Based Training}
In PSRO, a restricted game maintains a population of strategies (also known as policies, i.e., RL agents) and expands this population by introducing new strategies that respond to a specific distribution (mixture) of strategies within the existing population. 
Unlike self-play, which is a standard alternative approach in which a strategy is trained directly against itself, training a new strategy against a diverse set of strategies in the population, as in PSRO, can potentially enhance the robustness of the resulting strategy.
Consequently, PSRO can be classified as a variant of \newterm{population-based training}.
The textbook by Albrecht~et~al.~\shortcite{marl-book} provides an excellent explanation of PSRO from this viewpoint.

\subsection{Organization of the Survey}
In this survey, we discuss PSRO from the perspective of game-theoretic analysis, that is, how to conduct effective strategy exploration given a specific goal.
Although the performance of strategy exploration depends on the interplay between the chosen MSSs and ROs, existing literature predominantly focuses on setting either MSSs or ROs independently.
Therefore, we organize our discussion on research directions and corresponding PSRO variants by first discussing setting MSSs and ROs independently (Sections~\ref{sec: MSS} and \ref{sec: RO} respectively). 
We then, in Section~\ref{sec:joint}, discuss works that have investigated the joint choice of MSS-RO combination,
before moving on to discuss in Section~\ref{sec:evaluating} how best to evaluate the effectiveness of such choices for strategy exploration. 
We then discuss research on improving the efficiency of PSRO (Section~\ref{sec:efficiency}), explore applications of PSRO (Section~\ref{sec:applications}), implementations of PSRO (Section~\ref{sec:imp}), and conclude with open questions and future research directions (Section~\ref{sec:open}).

\section{Strategy Exploration via MSS}\label{sec: MSS}
In prior works, setting meta-strategy solvers was a primary way to control strategy exploration.
In this section, we discuss these works, their motivations, and their efficacy for strategy exploration.

\subsection{PSRO with Normal-Form Restricted Games}
In the standard PSRO framework, a restricted game is represented in normal form. 
While the simulations typically unfold through sequential observations and decisions over time, the restricted game abstracts away this temporal structure.

\subsubsection{Using Nash and its Variants as MSSs.}

In PSRO, the most common MSS target is NE, which can be computed by various game-theoretic methods based on the normal-form restricted game.
PSRO with NE is essentially DO with deep RL for computing (approximate) best responses.
Therefore, PSRO with NE inherits the convergence property of DO, given mild assumptions about the quality of the best responses:
In finite games, as long as beneficial deviations can always be found with non-zero probability, PSRO with NE as MSS target will converge to an NE given enough iterations.

\paragraph{\emph{The Overfitting Problem.}}
Despite its theoretical convergence guarantee, achieving exact convergence in large games is often unattainable due to constraints such as limited computational resources. 
Consequently, many prior works studying strategy exploration in large games revolve around the development of new algorithms that exhibit strong empirical performance (e.g., rapid convergence in terms of regret within a small number of PSRO iterations). 
These works found that a key problem that can prevent good empirical performance is \emph{overfitting}.
For strategy exploration, overfitting can arise in two distinct two ways.
First, strategy exploration may overfit to the NE of the restricted game.
Due to the limited information in the restricted game, the NE may not be an effective best-response target from a global view (i.e., using it, we may fail to generate non-trivial strategies for full-game play) while other best-response targets in the restricted game could be more effective.
The second form of overfitting relates to only capturing a specific equilibrium in general games (e.g., games with more than two players or general-sum games) without sufficiently exploring the whole strategy space. 
Note that the overfitting can be a problem for any solution concept; we discussed it in the context of NE specifically due to NE's prominence as the MSS target in the literature.

\paragraph{\emph{Regularization to Prevent Overfitting.}}
To address the overfitting problem, Lanctot~et~al.~\shortcite{lanctot2017unified} proposed an MSS, called Projected Replicator Dynamics (PRD), an adaptation of traditional replicator dynamics~\cite{taylor1978evolutionary}.
PRD ensures a probability lower bound for selecting each strategy in the restricted game, allowing the new best response to train against not only strategies in the equilibrium support but also those outside the support. 
PRD can be viewed as a form of \newterm{regularization} to 
prevent overfitting the response to a single exact NE of the restricted game.
Due to the diverse training targets, PRD also improves the stability of the new strategy.

Building on the concept of regularization, subsequent research has focused on designing MSSs that prevent overfitting by effectively regularizing the best-response target. 
For instance, Wang~et~al.~\shortcite{wang2019deep} proposed an MSS that combines NE with a uniform distribution as the best-response target, enabling the best response to an NE strategy mixed with exploration elements. 
Wright~et~al.~\shortcite{wright2019iterated} developed a history-aware approach with a best-response target mixed with previous targets.
Online double oracle~\cite{dinh2022online} integrated PSRO with online learning and used an online profile as the best-response target, which can be viewed as a form of regularization. 
Wang and Wellman~\shortcite{wang2023regularization} and Li~et~al.~\shortcite{li2023combining} employed quantal response equilibrium~\cite{mckelvey1995quantal,GempSLBA0TEK22} as an MSS, regularizing with bounded rationality.
Wang and Wellman~\shortcite{wang2023regularization} adopted an explicit view of regularization and introduced Regularized Replicator Dynamics (RRD), an MSS variant that truncates the NE search process in intermediate restricted games based on a regret criterion. 
Specifically, RRD computes the best-response target by running RD in the restricted game, stopping once the regret of the current profile with respect to the restricted game meets a specified regret threshold.
The regret criterion enables RRD to support direct control of the degree of regularization and can be adjusted to fit a specific game.

\subsubsection{MSSs Beyond Nash}
\paragraph{\emph{Rectified Nash}.}
Balduzzi~et~al.~\shortcite{balduzzi2019open} reformulated the strategy exploration problem as that of enlarging what they called the \emph{gamescape}, which describes the payoff space covered by the restricted game.
For symmetric two-player zero-sum games, they proposed \newterm{rectified Nash} as an MSS designed to expand the gamescape and enhance a diversity measure called effective diversity.
In rectified Nash, best responses are only applied to opponent's equilibrium strategies that the learning player defeats or ties with.

\paragraph{\emph{Minimum-Regret Constrained Profile.}}
One notable observation for PSRO with NE is that the full-game regret of the restricted-game NE, used as a measure for evaluating the performance of PSRO, does not decrease monotonically over PSRO iterations.
In the worst case, the full-game regrets will increase until the last iteration, when a full-game NE is found (one example can be found in the work of McAleer~et~al.~\shortcite{mcaleer1232021anytime}).
To address this issue, it was proposed to use \newterm{minimum-regret constrained profiles} (MRCP) \cite{Jordan10sw,wang2022evaluating} as an MSS.
An MRCP is the profile with minimum regret with respect to the full game\footnote{MRCP was called the \emph{least-exploitable restricted distribution} in the two-player zero-sum context by McAleer~et~al.~\shortcite{mcaleer1232021anytime}.}.
With MRCP as the MSS, the resulting PSRO variant is known as \emph{anytime} PSRO because regret monotonically decreases as the restricted game grows. 
Despite the difficulty of computing MRCP in general games, anytime PSRO leverages the properties of two-player zero-sum games and computes MRCP by regret minimization against a best response (RM-BR) \cite{johanson2012finding}.
In a further work \cite{mcaleer2022self}, anytime PSRO was extended by including not one but two strategies in the restricted game at each iteration, the first a best response to MRCP and the other a best response to the other player's latest strategy (i.e., the strategy added at the last PSRO iteration).
This modification was observed to improve the performance of anytime PSRO.
However, Wang et al.~\shortcite{wang2022evaluating} pointed out that MRCP regret will monotonically decrease for \emph{any} MSS since the concept of MRCP is well-defined in any restricted game, which suggests that using MRCP as an MSS in anytime PSRO is not justified purely by the desire to monotonically decrease MRCP regret.
We discuss the evaluation of strategy exploration further in Section~\ref{sec:evaluating}.

\paragraph{\emph{Correlated Equilibrium.}}
Apart from the issues discussed above, Marris~et~al.~\shortcite{marris2021multi} further argued that NE may not be an appropriate MSS for PSRO or even a solution concept in general-sum games due to its computational intractability. 
Therefore, they proposed utilizing correlated equilibrium~(CE) and coarse correlated equilibrium~(CCE) as MSSs, introducing a variant called Joint PSRO~(JPSRO). 
Since multiple (C)CE exist in a restricted game, they select the unique (C)CE that maximizes the Gini impurity.
Theoretical analysis demonstrated that JPSRO converges to a~(C)CE. 
Zhao~et~al.~\shortcite{zhao2023open} combined a diversity measure with~(C)CE and proposed a new MSS, called diverse (coarse) correlated equilibrium (DCCE). 
They showed the improved performance of PSRO with DCCE over JPSRO and many other PSRO variants.
Relatedly, Team-PSRO finds a \emph{team-maxmin equilibrium with coordination device}~\cite{celli2018computational}, an NE defined on the team level, in two-team zero-sum games~\cite{mcaleer2023teampsro}.

\paragraph{\emph{Game-Motivated MSSs.}}
In addition to the above well-established solution concepts, the literature on PSRO has also investigated other solution concepts, which originate from specific games but can be generally applied to other game settings. 
One example is the Risk-Averse Equilibrium (RAE) introduced by Slumbers~et~al.~\shortcite{slumbers2023game}, aimed at managing risk in multiagent systems. 
Specifically, RAE minimizes potential variance in rewards by accounting for the strategies of other players.
Another exampple is the work of Li~et~al.~\shortcite{li2023combining}, which proposed the Nash Bargaining Solution (NBS), a concept originating from the bargaining games, as an MSS. 
NBS can be computed by maximizing the product of players' utilities in the restricted game.

\paragraph{\emph{Automated MSS Design.}}
Distinct from prior works that design MSSs based on various solution concepts and heuristics, Feng~et~al.~\shortcite{feng2021neural} proposed Neural Auto-Curricula (NAC) based on meta-learning, which automates the design of MSSs in an end-to-end manner. 
Specifically, MSSs in NAC are parametrized by a neural network, which is trained by minimizing the regret of the meta-strategy in the resulting restricted games.
These restricted games are generated through PSRO with the current MSS (i.e., the current neural network) in games sampled from a game distribution.
With this training scheme, Feng~et~al.~\shortcite{feng2021neural} showed that NAC can learn an effective MSS for a family of games.
Automated MSS design with Auto-Curricula was also discussed by Yang~et~al.~\shortcite{yang2021diverse}, who highlighted the significance of including behavioral diversity in auto-curricula and presenting several challenges in designing such auto-curricula for successful real-world applications.
Another way to achieve automated MSS design is to select among existing MSSs to fit various games or different phases within a game adaptively.
For example, Li~et~al.~\shortcite{li2024self} applied hyperparameter optimization to learn weights among multiple MSSs and mixed the outputs of these MSSs based on the weights as the next best response target.

\subsection{PSRO with Alternative Game Forms}
\paragraph{\emph{Extensive-Form Games.}}
Instead of employing the normal-form representation, some PSRO variants use an extensive-form representation for the restricted game, offering a richer way to encompass temporal patterns in actions and information for underlying sequential games.
One such example is given by extensive-form DO (XDO) \cite{mcaleer2021xdo}.
The extensive-form restricted game tree in XDO again only contains a subset of players' strategies.
Similar to DO, NE is deployed as an MSS target, computed through Counterfactual Regret Minimization \cite{zinkevich2007regret}, and the best response computation will result in new actions at information states in the restricted game tree.
Note that when a new action is added to an information state, multiple strategies will be added.
So one iteration in XDO implicitly needs more simulation for profile evaluation than one DO iteration.
Periodic DO~\cite{pmlr-v202-tang23b} extends XDO by improving the stopping threshold of the restricted game solver. 

In a work by Konicki~et~al.~\shortcite{konicki2022exploiting}, the benefits of leveraging the extensive-form representation for the restricted game were further explored.
They showed that with an extensive-form representation in PSRO, the true game can be approximated more accurately than using a normal-form model constructed from the same amount of simulation data.
This accuracy improvement stems from the fact that the simulation data for modeling a chance node in extensive form can be reused while the modeling needs to be re-simulated every time for evaluating a profile in normal form.

\paragraph{\emph{Mean-Field Games.}}
Muller~et~al.~\shortcite{muller2022learning} and Wang and Wellman~\shortcite{wang2023empirical} adapted PSRO to mean-field games (MFGs).
Since the utility function for MFGs is not generally linear in the mean field, the restricted MFG cannot be represented explicitly.
Instead, Wang and Wellman~\shortcite{wang2023empirical} employed a game model learning approach \cite{sokota2019learning}, which is essentially a form of regression that learns a utility function over a restricted set of strategies and mean fields derived by these strategies.
They proved the existence of NE in the restricted MFG and the convergence of PSRO to NE in~MFGs.

\begin{table*}
    \centering
    \scalebox{1}{
    \setlength{\tabcolsep}{3pt}
    \begin{adjustbox}{width=1\textwidth}
        \begin{tabular}{c|c|cc|cc|cc}
            \toprule
            \multicolumn{1}{c}{\bf{Diversity Measure}}   &
            \multicolumn{1}{|c}{\bf{Concept}}   &
            \multicolumn{2}{|c}{\bf{Diversity Type}} &
            \multicolumn{2}{|c}{\bf{Enlargement Target}} &
            \multicolumn{2}{|c}{\bf{Compatible MSS}}\\
            \cmidrule(lr){3-4} 
            \cmidrule(lr){5-6}
            \cmidrule(lr){7-8}
             &   & Behavioral & Response & Gamescape & Policy Hull & Nash  & $\alpha$-Rank   \\
            \midrule
            Effective Diversity \cite{balduzzi2019open} & Rectified Nash strategy & & \checkmark & \checkmark & & \checkmark & \\
            \arrayrulecolor{lightgray}\hline
            Expected Cardinality \cite{perez2021modelling}  & Determinantal point processes & & \checkmark & \checkmark & & \checkmark & \checkmark \\
             \arrayrulecolor{lightgray}\hline
             Convex Hull Enlargement \cite{liu2021towards} & Euclidean projection  & & \checkmark & \checkmark & & \checkmark &  \\
            \arrayrulecolor{lightgray}\hline
             Occupancy Measure Mismatching \cite{liu2021towards}  & f-divergence, occupancy measure & \checkmark &  & \checkmark & & \checkmark &  \\
             \arrayrulecolor{lightgray}\hline
              Unified Diversity Measure \cite{liu2022unified} & Strategy feature, diversity kernel &  & \checkmark & \checkmark & & \checkmark & \checkmark \\
              \arrayrulecolor{lightgray}\hline
              Policy Space Diversity \cite{yao2023policy} & Bregman divergence, sequence-form  & \checkmark  & &  & \checkmark & \checkmark &  \\
            \arrayrulecolor{black}\bottomrule
        \end{tabular}
    \end{adjustbox} 
    }
    \caption{Specifications of promoting-diversity methods.}
    \label{table:diversity}
    \vspace{-0.3cm}
\end{table*}

\section{Strategy Exploration via RO}\label{sec: RO}
Besides setting MSSs, establishing ROs to guide strategy exploration in PSRO has also been explored in the literature.
One predominant way to design novel ROs is to include diversity measures in the standard RO, aimed at increasing the diversity of strategies in the restricted game. 
The importance of maintaining a diverse population of strategies has been demonstrated by many works in various domains~\cite{czarnecki2020real,zahavy2023diversifying}.

Specifically, Perez-Nieves~et~al.~\shortcite{perez2021modelling} proposed diverse PSRO, which incorporates expected cardinality, a diversity measure defined through a determinantal point process, into the standard RO. 
Liu~et~al.~\shortcite{liu2021towards} define behavioral diversity (BD) and response diversity (RD), measuring diversity on different scales. 
BD is defined based on the distance between action-state coverages given by different strategies, while RD measures the distance between the payoff vector induced by the new strategy and the current restricted game. 
Subsequently, Liu~et~al.~\shortcite{liu2022unified} proposed unified diversity measures to capture a variety of diversity metrics, which were later combined with the standard RO.
Yao~et~al.~\shortcite{yao2023policy} observed that the current diversity measures for enlarging the gamescape fail to connect to the quality of NE approximation.
They connected the RO to the quality of NE approximation by introducing population exploitability (PE), which is essentially the regret of MRCP \cite{wang2022evaluating}, to reflect the coverage of a policy hull (i.e., the convex combinations of strategies in the restricted game). 
They showed that a larger policy hull indicates lower PE. 
Therefore, they proposed an RO variant to enlarge the policy hull, aiming at promoting strategy exploration.
We list the specifications of different methods for enhancing diversity in Table~\ref{table:diversity}. 

Aside from diversity, Li~et~al.~\shortcite{li2023combining} deployed Monte Carlo tree search (MCTS) as the best-response oracle using different values (e.g., social welfare) to update values of nodes along the sample path in the back-propagation step of MCTS. 
The employment of different back-propagation values can also be viewed as modifications of the RO.

\section{Strategy Exploration via Joint MSS-RO}\label{sec:joint}
Generally speaking, MSSs and ROs have a coupled impact on strategy exploration.
In this section, we discuss prior works that jointly vary MSSs and ROs.

\paragraph{\emph{$\alpha$-Rank.}}
Muller~et~al.~\shortcite{muller2020generalized} proposed the adoption of {$\alpha$-Rank}~\cite{omidshafiei2019alpha} as the preferred solution concept due to its computational scalability and uniqueness in many-player general-sum games. 
To make PSRO with $\alpha$-Rank as MSS converge, they introduced the \newterm{preference-based best-response oracle}, which essentially returns a set of strategies that maximizes the probability mass under $\alpha$-Rank from the set of better responses to the current strategy profile.
With this MSS-RO combination, they proved that PSRO with $\alpha$-Rank converges to something that they call a \newterm{sink strongly-connected component}, which describes the distribution of strategies in long-term interactions.
Yang~et~al.~\shortcite{yang2019alpha} highlighted that as the number of players significantly increases, the Markov chain required in the computation of $\alpha$-Rank becomes prohibitively large, yielding a scalability problem for $\alpha$-Rank.
To handle this challenge, they developed an efficient implementation of $\alpha$-Rank based on DO and stochastic optimization.

\paragraph{\emph{Investigating the Joint Impact of MSSs and ROs.}}
An empirical study by Wang and Wellman~\shortcite{wang2024RO} explicitly investigated the coupling impact of MSSs and ROs in strategy exploration. 
This research experimented with many MSS-RO combinations with unique characteristics. 
Their experimental results underscore the pivotal role of ROs in steering strategy exploration towards desired objectives, such as higher social welfare.
Moreover, they showed that with a careful selection of MSS, the performance of strategy exploration can be further improved.

\section{Evaluating Strategy Exploration}\label{sec:evaluating}
In addition to designing novel strategy exploration algorithms, a significant effort has also been put into investigating methodological considerations in evaluating strategy exploration, and proposing and justifying new evaluation methods.
In PSRO, it may seem natural to employ the same MSS for both strategy generation and evaluation, as much of the prior works in PSRO exploration have done.
For example, if NE is used as the MSS for strategy generation, then the regret of NE of intermediate restricted games will be used as the performance measure at each iteration of PSRO.
Similarly, the regret of a uniform distribution over strategies will be the performance measure when the uniform distribution is employed as MSS, in which case PSRO reduces to fictitious play.

Wang and Wellman~\shortcite{wang2022evaluating} argued that this evaluation approach could yield a misleading conclusion about the performance of MSS-RO combinations. 
They highlighted that each MSS-RO combination essentially generates a distinct sequence of strategies, and thus the restricted game at any point reflects a distinct strategy space.
The comparisons of different MSS-RO combinations are across different strategy spaces, which may not be faithfully represented by a simple summary such as an interim solution.
Therefore, they proposed to use the regret of MRCP, where MRCP is the profile closest to the full-game NE (in regret) in the restricted game, as the evaluation metric for evaluating the performance of multiple MSS-RO combinations.
This means that the MSS employed for strategy generation is independent of the MSS (e.g., MRCP) used for evaluation, which should be fixed when comparing different restricted games, regardless of the MSS with which they are generated.

\section{Improvements in Training Efficiency}\label{sec:efficiency}
There are two components in PSRO that can be computationally demanding: best response computation and payoff simulation in the restricted game. 
To improve the efficiency of PSRO, various methods have been developed, addressing issues related to these two aspects.

\paragraph{\emph{Parallelization}.}
Leveraging parallelization, Lanctot~et~al.~\shortcite{lanctot2017unified} proposed the Deep Cognitive Hierarchy (DCH) model, which creates a training hierarchy where each player trains a best response strategy (with deep RL) against the NE of the restricted game with strategies at the same level or below it. 
This \emph{warm-starts} best-response training, and speeds up PSRO compared to training best responses from scratch.
Motivated by DCH, McAleer~et~al.~\shortcite{mcaleer2020pipeline} proposed Pipeline PSRO (P2SRO).
Similar to DCH, P2SRO initializes a bunch of strategies and assigns each strategy a level.
Then P2SRO warm-starts training each strategy in parallel against the NE of the restricted game involving strategies with lower levels, which accelerates the overall training of PSRO.

\paragraph{\emph{Sample Efficiency}.}
A distinctive characteristic of restricted games is that they are derived or estimated from simulation data.
To improve the sample efficiency of PSRO, Smith and Wellman~\shortcite{smith2023co} proposed to learn a full-game model about the game dynamics from the simulator concurrently with running PSRO, aimied at reducing the simulation cost by querying the full-game model. 
Zhou~et~al.~\shortcite{zhou2022efficient} developed an efficient PSRO (EPSRO) implementation for reducing the simulation cost of PSRO in two-player zero-sum games.
The key insight is that the simulation for the restricted game is only used for computing best-response target profiles. 
So as long as best-response target profiles can be computed in other ways (e.g., a uniform MSS does not need the evaluation of a restricted game), there is no need to maintain the complete restricted game, avoiding unnecessary simulations.

\paragraph{\emph{Transfer learning}.}
Transfer learning is a machine learning technique where a model trained on one task is repurposed for a different task. 
In PSRO, best-responding to different strategies can be viewed as such tasks and thus transfer learning can be applied to warm-start training new best responses.
One example leveraging this idea is NeuPL \cite {liu2022neupl}, which represents all strategies in the strategy set via a shared neural network.
NeuPL utilizes explicit parameter sharing for skill transfer, which was shown to effectively accelerate the adaptation to the opponent's meta-strategy.
Liu~et~al.~\shortcite{liu2022simplex} further generalized NeuPL by optimizing best-responses against mixed-strategy profiles randomly sampled from the current restricted game, offering approximate optimality against any mixture over a diverse set of strategies at test time.
Liu~et~al.~\shortcite{liu2024neural} combined NeuPL and JPSRO, enabling the transfer of knowledge in the computation of (C)CE with PSRO.
Smith~et~al.~\shortcite{smith2023strategic} also utilized transfer learning to reduce simulation costs in computing best responses. 
Their Mixed-Oracle method constructs a new best response by learning and maintaining best responses to the pure strategies of the opponent (represented as Q-value functions) and then mixing (Q-values) according to the meta-strategy.


\section{Applications}\label{sec:applications}
Game-theoretic analysis in PSRO relies on restricted games, which abstract away the underlying game structures.
This abstraction enables PSRO to solve a variety of games or address issues that can be formulated as a game, and yields numerous applications of PSRO in disparate domains.
Specifically, PSRO has been applied to specialized games, including security games~\cite{wang2019deep,wright2019iterated,fang2019integrate,tong2020finding,xu2021robust,cui2023macta}, bargaining games~\cite{li2023combining,wang2024RO}, Colonel Blotto games~\cite{an2023double}, Google Research Football environments~\cite{liu2021towards,song2024empirical}, chess~\cite{zahavy2023diversifying,li2023jiangjun}, Pursuit-Evasion games~\cite{li2023solving,li2024grasper}, auctions \cite{li2021evolution}, and mechanism design for sequential auctions~\cite{zhang2024computing}, which are among a class of the hardest extensive-form games to solve.

Moreover, PSRO applications over time have been extended to real-world domains including anti-jamming in satellite communication~\cite{zou2022equilibrium}, decision-making in beyond-visual-range air combat~\cite{ma2019cooperative}, solving the power imbalance in power system resilience~\cite{niu2021game}, and tackling the traveling salesman problem in combinatorial optimization~\cite{wang2021game}. 
Its utility is also evident in social network analysis for competitive influence maximization strategies~\cite{ansari2019competitive} and in developing defense strategies for election safety~\cite{yin2018optimal}.

Besides these real-world applications, PSRO (including DO) has also been applied for designing novel algorithms in domains that can be modeled as a game.
For example, the PSRO framework has facilitated developments in RL for robust policy discovery~\cite{liang2023game} and policy generalization~\cite{yang2022game}, multiagent RL evaluation~\cite{li2024meta}, value alignment in large language models~\cite{ma2023red}, public health services~\cite{killian2023robust}, combinatorial optimization~\cite{wang2024asp}, and the discovery of information in images~\cite{giboulot2023non}.

PSRO
has also had an impact on enhancing computer vision and structured prediction methodologies,
for example, via data augmentation for object detection~\cite{behpour2019ada}, active learning~\cite{behpour2019active}, semi-supervised multi-label classification~\cite{behpour2018arc}, and video tracking~\cite{fathony2018efficient}.
Furthermore, PSRO-type methods have been adapted to enhance the training of Generative Adversarial Networks (GAN)~\cite{oliehoek2019beyond,aung2022gan}.



\section{PSRO Implementations}
\label{sec:imp}

Two software libraries that include PSRO implementations are OpenSpiel~\cite{lanctot2019openspiel} and MALib~\cite{zhou2023malib}. 
Both serve as comprehensive toolsets including various games and algorithms for exploring general reinforcement learning and search or planning in games. 

\section{Open Research Questions}\label{sec:open}

The research directions outlined in the previous sections remain open-ended. 
Next, we describe several further research directions that we believe deserve the community's attention.

\paragraph{\emph{Scalability in the Number of Players}.}

In the standard PSRO framework, an empirical game is represented in normal form. 
As the number of players increases, the normal-form representation expands exponentially, leading to a significant increase in the cost of evaluating the empirical game. 
While this problem can be mitigated in certain special game types such as symmetric games, the practicality of applying PSRO diminishes when dealing with a very large number of players.
A potential remedy involves employing game model learning \cite{sokota2019learning}, essentially a utility function regression. 
The acquired utility function aims to extrapolate across the (empirical) strategy space, eliminating the need to evaluate each individual profile.

Another possible approach is to use \emph{polymatrix} game representations.
A polymatrix game is a many-player game  where each player can be represented as
a node in an interaction graph. 
On every edge in this
graph, there is a bimatrix game that the node plays with its neighbors, with the utility for each player being the sum of payoffs from these bimatrix games.
Polymatrix games have been used within equilibrium computation methods for many-player normal-form
games, due to their linear formulation.
In the work of both Govindan and Wilson~\shortcite{govindan2004computing} and Gemp at al.~\shortcite{GempSLBA0TEK22} a polymatrix approximation of 
the input game is maintained and updated (implicitly in the latter case).
There appears to be potential to combine or extract elements from these methods to build polymatrix-based 
empirical game models for PSRO-type methods.

\paragraph{\emph{A Completely Parallel PSRO Implementation}.}

Both best-response computation and empirical game simulation can benefit from parallelization. 
Firstly, best-response computations among players are parallelizable as they are independent. 
Secondly, the evaluation of strategy profiles in the empirical game can be parallelized. 
Additionally, since the evaluation of each strategy profile involves running simulations multiple times and averaging the results, these simulation runs can also be parallelized. 
Consequently, the PSRO framework is highly suitable for parallelization.
Despite this potential, there is currently no PSRO implementation that fully harnesses these three aspects.

\paragraph{\emph{Equilibrium Refinements}.}

As discussed in Section~\ref{sec: MSS}, PSRO can be adapted to calculating various solution concepts such as NE and CE. 
These solutions are often not unique, and the field of \newterm{equilibrium refinement} in game theory studies
restrictions of these concepts to make the resulting solutions more realistic and better predictions of how the games will/should be played~\cite{van2012refinements}.
A natural research direction is to design PSRO variants to compute equilibrium refinements.
For example, XDO, which uses an extensive-form representation, is a natural starting point to
develop equilibrium refinement methods within extensive-form games (e.g. finding subgame-perfect equilibria).
How can PSRO effectively compute these refinements by appropriately choosing the MSSs and ROs? 

\paragraph{\emph{Multiple Equilibria}.}
Another important question relates to the existence and computation of multiple equilibria, particularly within general-sum games. 
While current PSRO research primarily focuses on identifying a sample equilibrium, the capability of PSRO to compute multiple equilibria remains under-explored. 
In principle, one could enumerate all equilibria in the restricted game, and use all in separate calls to the best-response oracles.
In this way, we would be growing the restricted game in a way that encompasses \emph{all} strategic information collected so far. 
This approach requires maintaining a larger restricted game, which inherently demands increased computational resources and necessitates more efficient computational frameworks to manage the enlarged strategy space; it is an interesting research question to decide how to most effectively split the use of limited computational resources between responding to multiple equilibria in a single PSRO iteration, versus being able to complete more of those iterations.
Not all equilibria in the restricted game may be equally valuable for guiding strategy exploration. Can we predict this in advance?

\paragraph{\emph{Combining PSRO with Subgame Solving or CFR}.}
CFR-based and policy-gradient-based methods might perform better in games where mixing carefully at many decision nodes is crucial, providing more granular mixing at different game scales (e.g., information sets that define subgames).
In contrast, PSRO usually mixes only at the root of the game, limited to distributions over strategies in the restricted game, and might require many policies to effectively mix at information sets~\cite{mcaleer2021xdo}. 
Therefore, for game sections that do not require mixing such as complex control tasks, it would be more efficient to use the deep RL policies from PSRO or XDO, while decision nodes that require careful mixing should defer to a CFR-based or policy-gradient-based technique. 
Overall, a hierarchical PSRO structure that controls the granularity of the mixing might be needed.

Additionally, PSRO can be combined with subgame solving methods. 
Subgame solving methods, also called search methods, use extra compute at test time to analyze and improve a policy at a given information set without increasing exploitability. 
Some methods, such as depth-limited subgame solving~\cite{brown2018depth}, already iteratively expand the strategy space by adding best responses. 
It would be interesting to combine the best responses learned in the full game via PSRO with those in the subgame that are used for subgame solving. 

\paragraph{\emph{Automated Hyperparameter Tuning}.}
Hyperparameters exist universally in PSRO variants, including the lower bound of the probability of playing a strategy in Projected Replicator Dynamics \cite{lanctot2017unified}, the regret threshold in RRD \cite{wang2023regularization}, the probability of employing the uniform MSS \cite{wang2019deep}, and the weights for diversity regularizers \cite{perez2021modelling}.
These hyperparameters often require deliberate tuning to fit a specific game by running the same experiments multiple times for different hyperparameter choices, which can be costly.
Although some of the hyperparameters can be set heuristically (e.g., by employing a simple annealing scheme), the question of how to tune the hyperparameters automatically and efficiently remains open. 
Specifically, can we design an automatic adjustment scheme for these hyperparameters such that PSRO can be directly applied to any game without manual game-specific fine turning?
As a recent attempt, Li~et~al.~\shortcite{li2024self} proposed to learn a policy for hyperparameter optimization that can automatically select the optimal hyperparameter values in PSRO during game solving.
Moreover, they highlighted that a well-trained policy is generalizable to different games, thus reducing the cost of hyperparameter tuning when applying PSRO to various games.

\paragraph{\emph{PSRO with Large Language Models}.}
Recent achievements in large language models (LLMs) have drawn significant attention from various research communities.
In the study of multiagent systems, researchers have explored diverse methods for integrating LLMs with game-theoretic principles, each serving distinct objectives.
One research direction involves leveraging game-theoretic approaches to augment the capabilities of LLMs, such as enabling strategic responses in the presence of other agents \cite{gemp2024states}, or fostering alignment with human values \cite{ma2023red}. 
Conversely, LLMs can also be integrated into existing game-theoretic frameworks, expanding the range of applicable domains. 
Since PSRO provides a flexible game-theoretic framework, its potential to be combined with LLM in both ways should be noticed and investigated in future work.

\section*{Acknowledgements}
We thank Marc Lanctot, Karl Tuyls, Michael P. Wellman, Yaodong Yang, and their students, for their invaluable advice on improving the survey.

\section*{Call for Feedback}

We encourage and welcome suggestions and comments from our readers. 
Despite our best efforts, there is always a possibility of oversight, including missed citations or overlooked information. 
Therefore, we invite our audience to engage with the content critically and to provide feedback if they notice any discrepancies or omissions. 
Your insights and contributions are invaluable in ensuring the accuracy and integrity of our work.





\bibliographystyle{named_modified} 
\bibliography{ijcai24_modified}
\end{document}